\documentclass[letterpaper,twocolumn,prl,aps,superscriptaddress,amsmath,amssymb,floatfix]{revtex4-2}
\usepackage{mathptmx}
\usepackage[latin9]{inputenc}
\usepackage{color}
\usepackage{amsmath}
\usepackage{amssymb}
\usepackage{graphicx}
\usepackage{esint}
\usepackage{upgreek}  
\usepackage[unicode=true,
 bookmarks=true,bookmarksnumbered=false,bookmarksopen=false,
 breaklinks=false,pdfborder={0 0 1},backref=false,colorlinks=true]
 {hyperref}
\hypersetup{
 linkcolor=magenta,urlcolor=blue,citecolor=blue,pdfstartview={FitH},hyperfootnotes=false}

\makeatletter

\usepackage{textcomp}
\usepackage{epstopdf}

\pdfpageheight\paperheight
\pdfpagewidth\paperwidth

\@ifundefined{textcolor}{}{%
 \definecolor{BLACK}{gray}{0}
 \definecolor{WHITE}{gray}{1}
 \definecolor{RED}{rgb}{1,0,0}
 \definecolor{GREEN}{rgb}{0,1,0}
 \definecolor{BLUE}{rgb}{0,0,1}
 \definecolor{CYAN}{cmyk}{1,0,0,0}
 \definecolor{MAGENTA}{cmyk}{0,1,0,0}
 \definecolor{YELLOW}{cmyk}{0,0,1,0}
}

\usepackage{xcolor}\usepackage{soul}
\definecolor{blue}{rgb}{0,0,1}
\definecolor{red}{rgb}{1,0,0}
\definecolor{green}{rgb}{0,1,0}

\usepackage{soul}

\makeatother

\begin{document}

\title{Efficient single-atom transfer from an optical conveyor belt to a tightly confined optical tweezer}

\author{Lei Xu}
\affiliation{Key Laboratory of Quantum Information, University of Science and Technology of China, Hefei 230026, P. R. China.}
\affiliation{Anhui Province Key Laboratory of Quantum Network, University of Science and Technology of China, Hefei 230026, P. R. China}

\author{Ling-Xiao Wang}
\affiliation{Key Laboratory of Quantum Information, University of Science and Technology of China, Hefei 230026, P. R. China.}
\affiliation{Anhui Province Key Laboratory of Quantum Network, University of Science and Technology of China, Hefei 230026, P. R. China}

\author{Guang-Jie Chen}
\affiliation{Key Laboratory of Quantum Information, University of Science and Technology of China, Hefei 230026, P. R. China.}
\affiliation{Anhui Province Key Laboratory of Quantum Network, University of Science and Technology of China, Hefei 230026, P. R. China}

\author{Zhu-Bo Wang}
\affiliation{Key Laboratory of Quantum Information, University of Science and Technology of China, Hefei 230026, P. R. China.}
\affiliation{Anhui Province Key Laboratory of Quantum Network, University of Science and Technology of China, Hefei 230026, P. R. China}

\author{Xin-Biao Xu}
\affiliation{Key Laboratory of Quantum Information, University of Science and Technology of China, Hefei 230026, P. R. China.}
\affiliation{Anhui Province Key Laboratory of Quantum Network, University of Science and Technology of China, Hefei 230026, P. R. China}

\author{Guang-Can Guo}
\affiliation{Key Laboratory of Quantum Information, University of Science and Technology of China, Hefei 230026, P. R. China.}
\affiliation{Anhui Province Key Laboratory of Quantum Network, University of Science and Technology of China, Hefei 230026, P. R. China}
\affiliation{CAS Center For Excellence in Quantum Information and Quantum Physics,
University of Science and Technology of China, Hefei, Anhui 230026, P. R. China.}
\affiliation{Hefei National Laboratory, University of Science and Technology of China, Hefei 230088, China}

\author{Chang-Ling~Zou}
\email{clzou321@ustc.edu.cn}
\affiliation{Key Laboratory of Quantum Information, University of Science and Technology of China, Hefei 230026, P. R. China.}
\affiliation{Anhui Province Key Laboratory of Quantum Network, University of Science and Technology of China, Hefei 230026, P. R. China}
\affiliation{CAS Center For Excellence in Quantum Information and Quantum Physics,
University of Science and Technology of China, Hefei, Anhui 230026, P. R. China.}
\affiliation{Hefei National Laboratory, University of Science and Technology of China, Hefei 230088, China}

\author{Guo-Yong Xiang}
\email{gyxiang@ustc.edu.cn}
\affiliation{Key Laboratory of Quantum Information, University of Science and Technology of China, Hefei 230026, P. R. China.}
\affiliation{Anhui Province Key Laboratory of Quantum Network, University of Science and Technology of China, Hefei 230026, P. R. China}
\affiliation{CAS Center For Excellence in Quantum Information and Quantum Physics,
University of Science and Technology of China, Hefei, Anhui 230026, P. R. China.}
\affiliation{Hefei National Laboratory, University of Science and Technology of China, Hefei 230088, China}

\date{\today}

\begin{abstract}
Efficient loading of single atoms into tightly confined traps is crucial for advancing quantum information processing and exploring atom-photon interactions. However, directly loading atoms from a magneto-optical trap (MOT) into static tweezers in cavity-based systems and hybrid atom-photon interfaces remains a challenge. Here, we demonstrate atom loading in a tightly confined optical tweezer $0.6\,\mathrm{mm}$ away from MOT by an optical conveyor belt. By employing real-time feedback control of the atom number in the overlapping region between the conveyor belt and the tweezer, we enhance a single-atom loading probability to 77.6\%. Our technique offers a versatile solution for deterministic single-atom loading in various experimental settings and paves the way for diverse applications based on hybrid photonic-atom structures. 
\end{abstract}
\maketitle

\section{Introduction}

Single atoms trapped in optical tweezers, i.e., tightly confined optical dipole traps, have emerged as a powerful platform for a variety of quantum technologies, including quantum information processing~\cite{darquie2005Controlled,fortier2007Deterministic, saffman2010Quantum,graham2022Multiqubit}, quantum simulation~\cite{ebadi2021Quantum,kaubruegger2019Variational,browaeys2020Manybodya}, and quantum sensing~\cite{zektzer2021Nanoscale,sebbag2021ChipScale}. The ability to individually control and manipulate the external degrees of freedom of atoms in tweezers enables the preparation of well-defined quantum states encoded in their long-coherence internal states~\cite{kaufman2012Cooling,thompson2013Coherence,meng2020Imaging,tian2024Resolved}, facilitating the realization of high-fidelity quantum gates between atoms and photons. In particular, atoms can be trapped by dipole traps inside an optical cavity or in the vicinity of a photonic chip for an efficient atom-photon interface. To load single atoms into tweezers, a common approach is to overlap the tweezers with a cold atom cloud produced by a magneto-optical trap (MOT), resulting in probabilistic loading of single atoms due to light-assisted atomic collisions~\cite{schlosser2001Subpoissonian,schlosser2002Collisional}. By detecting the presence of single atoms in each tweezer, subsequent feedback rearrangement of the tweezer array can then be used to create defect-free arrays of single atoms~\cite{barredo2016Atombyatom,endres2016Atombyatom}.

However, this approach faces challenges when applied to specific experimental configurations. For dipole traps in Fabry-Perot cavities~\cite{fortier2007Deterministic,liu2023Realization,deist2022Superresolution,nussmann2005Submicron,reimann2015CavityModified}, it is crucial to have a static trap inside the cavity, which prevents overlapping with the MOT and limits the ability to rearrange the atoms~\cite{wang2025PurcellEnhanced,wang2025Cavity}. Similarly, in hybrid quantum systems where dipole traps are produced by photonic microstructures near the chip surface, overlapping the traps with the atom cloud and manipulating the tweezers can be difficult~\cite{kim2019Trapping,zhou2023Coupling,Xu2023,bouscal2024Systematic,Chen2025}. Previously, free-falling MOTs were used to load the tight trap randomly, resulting in a very low loading probability~\cite{aoki2006Observation,scheucher2016Quantum,shomroni2014Alloptical}. In these cases, an efficient cold atom pipeline that can deliver atoms from the MOT to the tweezers and achieve near-deterministic loading of single atoms into static traps is of great importance~\cite{liu2023Proposal,chiu2025Continuous}.

In this work, we experimentally demonstrate an approach for efficiently loading single atoms into a static optical tweezer from an optical conveyor belt. By employing real-time feedback control, we achieve a single-atom loading efficiency of $77.6\%$ by controlling the transportation of atoms within an optical conveyor belt. Furthermore, we investigate the influence of the conveyor belt trap depth on the atom loss rate and the ability to distinguish single atoms, providing insight into the optimal conditions for efficient single-atom loading. This work extends the application of single-atom tweezers by enabling high loading probabilities and efficient atom-photon interfaces, paving the way for advances in quantum applications with single atoms.

\begin{figure*}[ht]
\centering{\includegraphics[width=\linewidth]{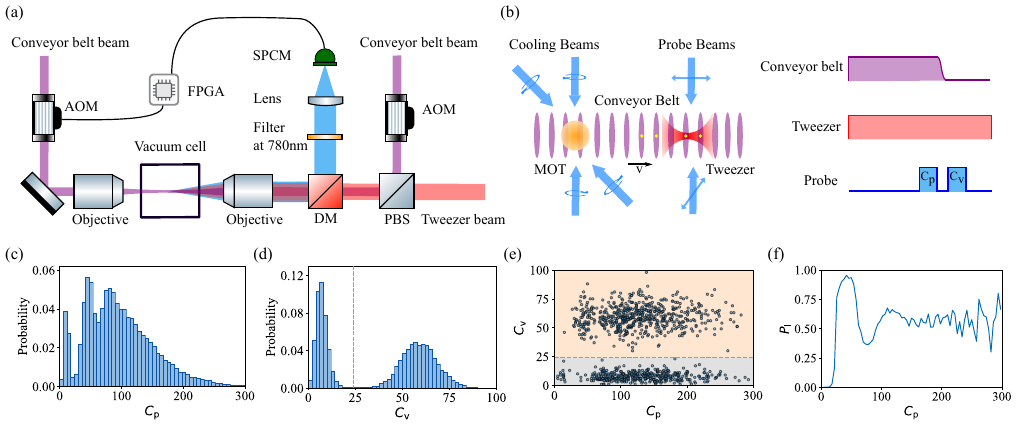}}
\caption{Single-atom transfer from an optical conveyor belt to a tightly confined optical tweezer. (a) Schematic of the experimental setup for transferring single atoms from an optical conveyor belt trap (purple) into a tightly focused optical tweezer trap (red). Both traps are created using Gaussian beams at a wavelength $\lambda$ of $852\,\mathrm{nm}$, focused through a shared objective lens (numerical aperture 0.28). The conveyor belt trap has a waist $w_{\mathrm{c}}$ of $10\,\mu\mathrm{m}$, and the tweezer trap has a waist $w_{\mathrm{t}}$ of $2\,\mu\mathrm{m}$. The trap beams are combined using a polarizing beam splitter (PBS). Fluorescence from atoms is isolated from background light via a dichroic mirror (DM) and detected by a single-photon counting module (SPCM). Real-time signal processing is performed with a field-programmable gate array (FPGA), which controls acousto-optic modulators (AOMs) to drive the conveyor belt movement. (b) Illustration of the atom transfer process. Cold atoms are loaded from a magneto-optical trap (MOT) and transported to the overlap region of the conveyor belt and tweezer traps. The conveyor belt is then ramped down to zero, leaving a single atom confined within the tweezer trap. Fluorescence counts $C_{\mathrm{p}}$ (before ramp-down) and $C_{\mathrm{v}}$ (after ramp-down) are measured using two linearly polarized probe beams. (c, d) Histogram of fluorescence counts $C_{\mathrm{p}}$ and $C_{\mathrm{v}}$ measured over 50 ms exposure periods with 20000 repeated measurements, before and after the conveyor belt is ramped down, respectively. The dashed lines represent the count thresholds used for identifying a single atom in the tweezer. (e) Scatter plot of $C_{\mathrm{p}}$ vs. $C_{\mathrm{v}}$ for individual experimental trials. Trials where $C_{\mathrm{v}}$ exceeds the threshold (dashed line) are classified as successful single-atom loading. (f) Single-atom loading probability $P_{\mathrm{l}}$ as a function of the initial fluorescence count $C_{\mathrm{p}}$.}
\label{fig1}
\end{figure*}

\section{Experimental setup}

Figure~\ref{fig1}(a) shows the experimental setup for loading a single $^{87}$Rb atom into an optical tweezer from an optical conveyor belt. The conveyor belt trap is created by two counter-propagating Gaussian beams, each with a wavelength $\lambda$ of $852 \, \text{nm}$, through a lens (numerical aperture 0.28). This setup achieves a beam waist $w_\mathrm{c}$ of $10 \,\mathrm{\mu m} $. The tweezer trap is formed by a tightly focused Gaussian beam with a waist $w_\mathrm{t}$ of $2 \,\mu \text{m} $. Both beams, the conveyor belt and the tweezer, originate from the same 852\,nm laser source. They are subsequently combined via a polarizing beam splitter (PBS) and directed into the vacuum cell.

The one-dimensional trap lattice of the conveyor belt can be moved by varying the frequency difference $\delta$ between the beams forming the conveyor belt using two phase-locked acousto-optic modulators (AOMs), achieving a velocity of $v=\frac{1}{2}\lambda \delta$~\cite{kuhr2001Deterministic}. The fluorescence photons emitted by the trapped atoms are probed at the cycling D2 transition with a wavelength of 780 \,nm. These photons are separated from the trapping beams using a dichroic mirror (DM), then coupled into a single-mode fiber and detected by a single-photon counting module (SPCM). The tweezer is precisely aligned to ensure that the foci of both the 852\,nm dipole laser and the 780\,nm atom fluorescence coincide, thus enabling the detection of atoms trapped within the overlapping region of the two traps. A field-programmable gate array (FPGA) processes the photon counts to enable real-time feedback control of the conveyor belt lattice movement.

We first evaluate the single-atom loading probability by directly overlapping the two trapping beams, as illustrated in Fig.~\ref{fig1}(b). A cold $^{87}$Rb atom cloud, prepared using a standard magneto-optical trap (MOT), has a radius of about $100\,\mathrm{\mu m}$ and is initially positioned about $600\,\mathrm{\mu m}$ away from the optical tweezer. Atoms are loaded into the conveyor belt lattice and transported toward the tweezer trap. The tweezer trap depth is set to $U_{\mathrm{t}}=1\,\mathrm{mK}$, while the conveyor belt trap depth is $U_{\mathrm{c}}=0.3\,\mathrm{mK}$. Atoms in the overlapping trap region are probed using a pair of linearly polarized counterpropagating 780 nm laser beams, which also provide polarization gradient cooling (PGC). Figure~\ref{fig1}(c) shows the fluorescence counts $C_{\mathrm{p}}$ in the conveyor belt trap under a 50 ms exposure time. Clear multi-atom fluorescence signals are observed, with background and single-atom peaks well-resolved. However, atom loss in the conveyor belt trap reduces fluorescence counts, washing out discrete peaks for two or more atoms. To transfer atoms into the tweezer trap, we adiabatically ramp down the conveyor belt trap to zero within 2 ms and apply a second 50 ms probe pulse for parity projection, resulting in either one or zero atoms in our tweezer. Figure~\ref{fig1}(d) shows the fluorescence counts $C_{\mathrm{v}}$  over a 50 ms exposure time in the tweezer. Distinct separation between atom fluorescence and background counts confirms single-atom loading events. Multi-atom loading events are suppressed due to light-assisted collisions. The dashed line indicates the fluorescence threshold (24 counts per 50\,ms) used to identify single-atom loading events, resulting in a measured single-atom loading probability $P_{\mathrm{l}}$ of $57.5\%$.

Then we analyze the relation of fluorescence counts $C_{\mathrm{p}}$ and $C_{\mathrm{v}}$ before and after the conveyor belt is ramped down. Figure~\ref{fig1}(e) compares $C_{\mathrm{p}}$ and $C_{\mathrm{v}}$ for each experimental trial. Trials with $C_{\mathrm{v}}$ above the dashed line correspond to single-atom loading into the tweezer trap. We evaluate the single-atom loading probability $P_{\mathrm{l}}$ as a function of $C_{\mathrm{p}}$, as shown in Fig.~\ref{fig1}(f). Nearly 95\% single-atom loading probability is achieved when the photon counts $C_{\mathrm{p}}$ are around 40, indicating efficient single-atom transfer. In contrast, the loading probability drops below 40\% as $C_{\mathrm{p}}$ increases to around 80, suggesting that two atoms are loaded into the tweezer but are subsequently lost due to light-assisted collisions. At higher fluorescence levels, the loading probability stabilizes near 50\%, consistent with sub-Poissonian atom number statistics in a tightly confined trap. Increased fluctuations in loading probability at higher photon counts are attributed to reduced sampling numbers. These results demonstrate that the collected fluorescence counts $C_{\mathrm{p}}$ serve as a reliable indicator for identifying the presence of a single atom in the overlapping trap region with high confidence.

\section{Results}

\begin{figure}[hbt]
\center{\includegraphics[scale=1]{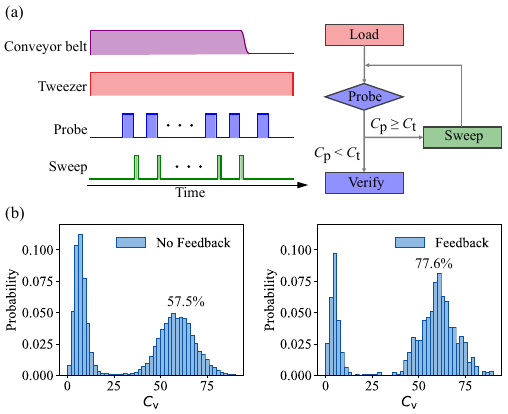}}
\caption{Feedback control for the single-atom transfer from an optical conveyor belt to a tightly confined optical tweezer. (a) Time sequence of the feedback control protocol. Fluorescence counts $C_{\mathrm{p}}$ are measured to infer the atom number  within the overlapping region of the conveyor belt and tweezer traps. An FPGA compares $C_{\mathrm{p}}$ to a predefined threshold $C_{\mathrm{t}}$ to determine whether feedback is necessary. If so, a 10 ms frequency-sweeping pulse is applied to shift the conveyor lattice. If not, the conveyor belt is ramped down to verify the single-atom transfer into the tweezer trap. (b) Histograms of fluorescence counts recorded in the tweezer trap without and with feedback control. The implementation of feedback increases the single atom loading probability from 57.5\% to 77.6\%.}
\label{fig2}
\end{figure}

The key strategy for improving single-atom loading probability is to deterministically control the number of atoms transferred into the optical tweezer trap.  As illustrated in Fig.~\ref{fig2}(a), fluorescence counts are measured over a 50 ms exposure time to infer the number of atoms within the overlap region of the two traps. Based on the detected signal $C_{\mathrm{p}}$, an FPGA determines whether feedback is needed by comparing the signal to a pre-set threshold $C_{\mathrm{t}}$. The FPGA then adjusts the frequency difference $\delta$ applied to the AOMs to shift the conveyor belt accordingly. If multiple atoms are detected, i.e., $C_{\mathrm{p}} \geq C_{\mathrm{t}}$, a 10 ms frequency-sweeping pulse is applied to shift the lattice by one step, modifying the local atom number. This procedure is repeated until either a single atom is detected or a maximum feedback duration of 1 s is reached. Once a valid single-atom signal is identified, i.e., $C_{\mathrm{p}} < C_{\mathrm{t}}$, the conveyor belt is ramped down to zero, completing the transfer into the tightly confined tweezer trap. Figure~\ref{fig2}(b) presents histograms of fluorescence counts in the tweezer trap, comparing results obtained without and with feedback control. Implementing feedback increases the single-atom loading probability from 57.5\% to 77.6\%. In this experiment, the lattice is displaced by $3.75\,\mu\mathrm{m}$ per step. The tweezer trap depth $U_{\mathrm{t}}$ is fixed at $1.2\, \text{mK}$, the conveyor belt trap depth $U_{\mathrm{c}}$ is set to $0.62\, \text{mK}$, and the threshold count $C_{\mathrm{t}}$ is set to 40 to optimize performance.

\begin{figure}[hbt]
\center{\includegraphics[scale=1]{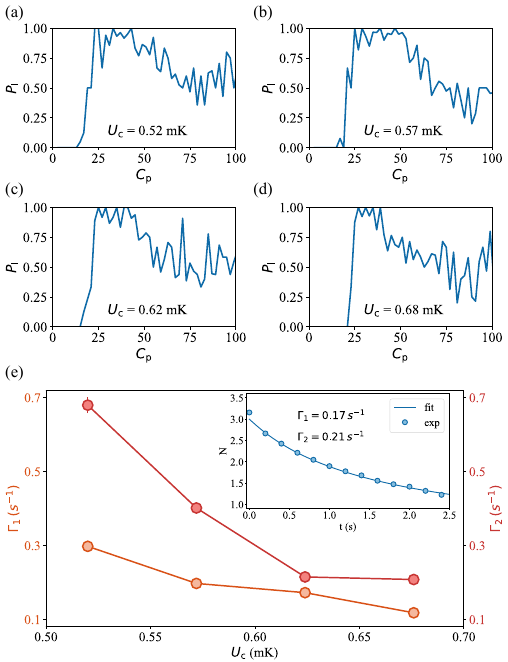}}
\caption{Influence of conveyor belt trap depth. (a-d) Single-atom loading probability $P_{\mathrm{l}}$ as a function of the fluorescence count $C_{\mathrm{p}}$ for different conveyor belt trap depths $U_{\mathrm{c}}$. The tweezer trap depth $U_{\mathrm{t}}$ is fixed at $1.2\, \text{mK}$. Increasing $U_{\mathrm{c}}$ results in a reduction of the fluorescence counts window size. (e) Measured atom loss rate as a function of conveyor belt trap depth $U_{\mathrm{c}}$. Error bars represent the 95\% confidence interval of the fitted atom loss rate. The inset shows an example of the averaged atom number $N$ versus probe time $t$, with fitted single-atom loss rate of $ 0.17 \,s^{-1}$ and two-atom loss rate of $ 0.21 \,s^{-1}$.
}
\label{fig3}
\end{figure}

Next, we vary experimental parameters to optimize the single-atom loading probability. A large fluorescence counts window, as illustrated in Fig.~\ref{fig1}(f), is crucial for accurately distinguishing situations where only a single atom remains within the overlapping region. In addition, minimizing atom loss during the feedback control procedure is essential to allow multiple feedback cycles and reliably load a single atom. Figures~\ref{fig3}(a)-(d) show the single-atom loading probability $P_{\mathrm{l}}$ as a function of the initial fluorescence counts $C_{\mathrm{p}}$ for different trap depths $U_{\mathrm{c}}$. The tweezer trap depth $U_{\mathrm{t}}$ is fixed at $1.2\, \text{mK}$. The conveyor belt trap depths $U_{\mathrm{c}}$ are set to $0.52\, \text{mK}$, $0.57\, \text{mK}$, $0.62\, \text{mK}$, and $0.68\, \text{mK}$ in (a-d), respectively. An increase in $U_{\mathrm{c}}$ leads to a reduction in the fluorescence counts window. This is attributed to the increased ac Stark shift, which modifies the fluorescence scattering rate of atoms trapped at different antinodes of the conveyor belt trap. Consequently, this variation disrupts the ability to accurately determine the presence of a single atom within the overlapping trap region. 

By fitting the evolution of the atom number in the trap, as shown in the inset of Fig.~\ref{fig3}(e), we can extract the single-atom and two-atom loss rates $\Gamma_1$ and  $\Gamma_2$ ~\cite{schlosser2002Collisional,wang2023Controllable}, which are described by the equation
\begin{equation}
\frac{\mathrm{d}N}{\mathrm{d}t} = -\Gamma_1 N -\Gamma_2 N(N-1),
\label{eq1}
\end{equation}
where $N$ is the atom number that is determined by the identified fluorescence step of a single atom. Then, we can examine the impact of the conveyor belt trap depth $U_{\mathrm{c}}$ on the atom loss rate, as shown in Fig.~\ref{fig3}(e). In Ref.~\cite{wang2023Controllable}, it is reported that a modulated tight optical dipole trap can modify both $\Gamma_1$ and $\Gamma_2$. Similarly, in our experiment, increasing the conveyor belt trap depth reduces $\Gamma_1$ and $\Gamma_2$. For example, when the conveyor belt trap depth is set to $U_{\mathrm{c}} = 0.62\,\text{mK}$, the single-atom and two-atom loss rate is reduced to 0.17 $ \text{s}^{-1}$ and 0.21 $ \text{s}^{-1}$. 

\begin{figure}[hbt]
\center{\includegraphics[scale=1]{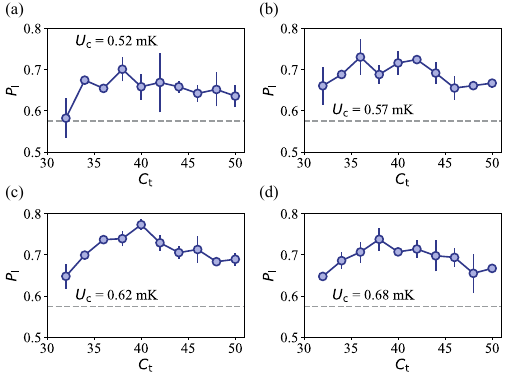}}
\caption{Single-atom loading probability $P_{\mathrm{l}}$ as a function of threshold counts $C_{\mathrm{t}}$ for different conveyor belt trap depths $U_{\mathrm{c}}$. The tweezer trap depth $U_{\mathrm{t}}$ is fixed at $1.2\, \text{mK}$. Each measurement is repeated 250 times. Error bars represent the $1\sigma$ standard error. The optimal loading probability of 0.776 (0.012) is achieved when the conveyor belt trap depth is set to $0.62\,\text{mK}$ and threshold count $C_{\mathrm{t}}$ is set to 40.
}
\label{fig4}
\end{figure}

Based on these experimental results, increasing the conveyor belt trap depth reduces the fluorescence counts window while decreasing the atom loss rate. Therefore, achieving optimal single-atom loading probability requires balancing the reduction in the atom loss rate with maintaining a sufficiently large fluorescence counts window to accurately distinguish single atoms within the overlapping trap region. Figure~\ref{fig4} shows the single-atom loading probability versus the  threshold counts  for different conveyor belt trap depths. To optimize the performance, threshold fluorescence counts $C_{\mathrm{t}}$ are adjusted for each trap depth. The optimal loading probability of 77.6\% is achieved when the conveyor belt trap depth is set to $0.62\,\text{mK}$ and the threshold count $C_{\mathrm{t}}$ is set to 40, as shown in Fig.\ref{fig4} (c).

\section{Discussion and conclusion}

Finally, we analyze the limitations of our method for deterministic loading of a single atom from an optical conveyor belt. Several factors can lead to the failure of our feedback control strategy. First, there is an approximately $8.6\%$ probability that no atoms are initially loaded into the region where the optical tweezer and conveyor belt traps overlap. Second, single-atom loss can occur when ramping down the conveyor belt beam. To quantify this loss, separate experiments are conducted where single atoms are prepared in the conveyor belt trap, and the transfer probabilities to the optical tweezer are measured. These measurements reveal a single-atom loss probability of approximately $4\%$, primarily caused by the non-adiabatic reduction of the conveyor belt trap depth and misalignment between the two trap beams. Finally, fluctuations in single-atom fluorescence photon counts can result in counts exceeding the predefined counts threshold, leading to continued feedback processes that displace the single atom from the overlap region.

In summary, we have demonstrated an efficient method for single-atom loading into an optical tweezer from an optical conveyor belt. Using feedback control to change the atom number transferred, a single atom is deterministically loaded into the tweezer with an improved loading probability of $77.6\,\%$. The feedback process requires an average duration of 260 ms in the current experiments, which consist of four feedback loops. This duration can be further reduced by decreasing the fluorescence detection time and the conveyor belt frequency-sweeping time. Importantly, our feedback control method introduces minimal additional time to each experimental cycle, enabling a high repetition rate for single-atom preparation. By combining the long-range atom transport capability of the optical conveyor belt, this method is applicable for single-atom loading into surface traps on nanophotonic devices. Atoms confined at the antinodes of the conveyor belt trap along its axis can be repeatedly loaded into the surface trap. Furthermore, the loading probability from the optical conveyor belt could be enhanced by implementing feedback control to identify successful single-atom loading events, potentially verified by probing the atom's coupling to nanophotonic devices. Ultimately, this approach enables deterministic single-atom loading on integrated photonic devices and facilitates controllable interactions between single atoms and single photons.

\section{Acknowledgments}

\begin{acknowledgments}
This work was funded by the National Key R \& D Program (Grant No. 2021YFA1402004), the National Natural Science Foundation of China (Grant Nos.~12134014, U21A6006, 92265210, and 12293053). This work was also supported by the Fundamental Research Funds for the Central Universities, USTC Research Funds of the Double First-Class Initiative. The numerical calculations in this paper have been done on the supercomputing system in the Supercomputing Center of University of Science and Technology of China. This work was partially carried out at the USTC Center for Micro and Nanoscale Research and Fabrication.
\end{acknowledgments}

\end{document}